\documentclass[fleqn,usenatbib]{mnras}

\usepackage{newtxtext,newtxmath}

\usepackage[T1]{fontenc}

\DeclareRobustCommand{\VAN}[3]{#2}
\let\VANthebibliography\thebibliography
\def\thebibliography{\DeclareRobustCommand{\VAN}[3]{##3}\VANthebibliography}

\usepackage{graphicx}	

\usepackage{savesym}
\savesymbol{Bbbk}
\usepackage{caption}
\usepackage{amsmath}	
\usepackage{amssymb}	
\restoresymbol{Bbbk1}{Bbbk}
\usepackage[dvipsnames]{xcolor}

\title[Magnetic fields and outflows in CB 54]{Magnetic fields and outflows in the large Bok globule CB 54}

\author[K. Pattle et al.]{
Kate Pattle,$^{1,2}$\thanks{E-mail: k.pattle@ucl.ac.uk (KP)}
Shih-Ping Lai,$^{3,4}$, Sarah Sadavoy,$^{5}$ Simon Coud\'{e}$^{6}$, Sebastian Wolf$^{7}$, Ray Furuya,$^{8}$\newauthor Woojin Kwon,$^{9,10}$ Chang Won Lee$^{11,12}$ and Niko Zielinski$^{7}$
\\
$^{1}$Department of Physics and Astronomy, University College London, Gower Street, London WC1E 6BT, United Kingdom \\
$^{2}$Centre for Astronomy, Department of Physics, National University of Ireland Galway, University Road, Galway H91 TK33, Ireland\\
$^{3}$Institute of Astronomy and Department of Physics, National Tsing Hua University, No. 101, Section 2, Guangfu Road, Hsinchu 30013, Taiwan\\
$^{4}$Academia Sinica Institute of Astronomy and Astrophysics, No. 1, Section 4., Roosevelt Road, Taipei 10617, Taiwan \\
$^{5}$Department for Physics, Engineering Physics and Astrophysics, Queen's University, Kingston, ON, K7L 3N6, Canada \\
$^{6}$SOFIA Science Center, Universities Space Research Association, NASA Ames Research Center, Moffett Field, California 94035, USA\\
$^{7}$Institut f\"ur Theoretische Physik und Astrophysik, Christian-Albrechts-Universit\"at zu Kiel, Leibnizstr. 15, 24118 Kiel, Germany\\
$^{8}$Institute of Liberal Arts and Sciences, Tokushima University, Minami Jousanajima-machi 1-1, Tokushima 770-8502, Japan \\
$^{9}$Department of Earth Science Education, Seoul National University, 1 Gwanak-ro, Gwanak-gu, Seoul 08826, Republic of Korea \\
$^{10}$SNU Astronomy Research Center, Seoul National University, 1 Gwanak-ro, Gwanak-gu, Seoul 08826, Republic of Korea \\
$^{11}$Korea Astronomy and Space Science Institute, 776 Daedeokdae-ro, Yuseong-gu, Daejeon 34055, Republic of Korea \\
$^{12}$University of Science and Technology, Korea, 217 Gajeong-ro, Yuseong-gu, Daejeon 34113, Republic of Korea 
}

\date{Accepted XXX. Received YYY; in original form ZZZ}

\pubyear{2022}

\begin{document}
\label{firstpage}
\pagerange{\pageref{firstpage}--\pageref{lastpage}}
\maketitle

\begin{abstract}
We have observed the large Bok globule CB 54 in 850$\mu$m polarised light using the POL-2 polarimeter on the James Clerk Maxwell Telescope (JCMT).  We find that the magnetic field in the periphery of the globule shows significant, ordered deviation from the mean field direction in the globule centre.  This deviation appears to correspond with the extended but relatively weak $^{12}$CO outflow emanating from the Class 0 sources at the centre of the globule.  Energetics analysis suggests that if the outflow is reshaping the magnetic field in the globule's periphery, then {we can place an upper limit of $<27\,\mu$G on the magnetic field strength in the globule's periphery.} 
Comparison with archival \textit{Planck} and CARMA measurements shows that the field in the centre of the globule is consistent over several orders of magnitude in size scale, {and} oriented parallel to the density structure in the region in projection.  We thus hypothesise that while non-thermal motions in the region may be sub-Alfv\'enic, the magnetic field is subdominant to gravity {over a wide range of size scales}.  Our results suggest that even a {relatively} weak outflow {may be able to} significantly reshape {magnetic} fields in star-forming regions on scales $> 0.1$\,pc.  
\end{abstract}

\begin{keywords}
stars: formation -- ISM: magnetic fields -- submillimetre:ISM
\end{keywords}



\section{Introduction}
\label{sec:introduction}

Outflows from protostellar systems play an important role in the dynamics of the molecular clouds from which they form, providing instantaneous feedback on the cloud as they eject angular momentum and kinetic energy from the forming stellar system \citep{bally2016}.  Molecular clouds, like the lower-density interstellar medium (ISM), are threaded by magnetic fields on all size scales \citep[e.g.][]{crutcher2012}.  Outflows and magnetic fields in combination may play a significant role in setting the star formation efficiency of both cores \citep[e.g.][]{offner2017} and clouds \citep[e.g.][]{krumholz2019}.

Interferometric observations of large samples of protostellar sources have found that magnetic fields and outflows are typically randomly aligned, with the possibility of a well-aligned subset, on interferometric (1000s of au) scales \citep{hull2014,hull2019}, while a comparable single-dish study found outflows to be misaligned with respect to magnetic fields by an average of $50^{\circ}\pm 15^{\circ}$ in 3D, but did not rule out random orientation \citep{yen2021}.
Recent observations have shown that in some cases, dust polarisation around young protostars appears to trace their outflow cavity walls, suggesting compression and rearrangement of magnetic fields by outflow feedback \citep{hull2020, lyo2021}.  However, the physical scales over which and timescales on which this rearrangement can occur are not well-established, and neither is the range of magnetic environments in which this rearrangement of the field can take place.

Bok globules \citep{bok1947} are isolated clumps of molecular gas, typically containing a few tens of solar masses within a diameter of a few tenths of a parsec \citep{launhardt2010}.  They are therefore a relatively simple environment in which the low-mass star formation process can be studied.

CB 54 \citep{clemens1988}, also known as LBN 1042 \citep{lynds1965}, is a large ($\sim 0.5\,$pc diameter) globule at a distance of $\sim 1.5$\,kpc which is associated with the Vela OB1 complex \citep{ciardi2007,sen2021}. CB 54 is an example of a forming group of low-mass stars \citep{ciardi2007}, and so whether it should be described as a Bok globule or as a small cloud is arguable.  As the maximum mass of a Bok globule is not well-defined, and as CB 54 is an isolated, {apparently spheroidal} cloud {(approximately circular in projection)} and is consistently referred to as a Bok globule in the literature, we continue to define it as such in this work.  

CB 54 contains five known point sources: the NIR-bright sources CB54YC1-I and -II \citep{yun1996}, and the deeply-embedded sources MIR-a, -b and -c \citep{ciardi2007}.  CB54YC1-II is established as a Class I source \citep{yun1996}; the classification of CB54YC1-I is more uncertain, with possibilities including its being a highly extinguished embedded A or B star, a background G or F giant, or a very highly-inclined embedded protostar \citep{ciardi2007}.  The three remaining sources, MIR-a, -b and -c, are deeply-embedded Class 0 sources forming in close proximity to one another \citep{ciardi2007}, coincident with the IRAS source PSC 07020–1618 and with a peak in submillimetre emission \citep{launhardt2010}. At least one of these sources is driving a small-scale molecular outflow oriented 108$^{\circ}$ E of N \citep{hull2014}.  A larger-scale outflow from CB 54, oriented broadly NE/SW ($\sim45^{\circ}$ E of N), was mapped by \citet{yun1994}.  \citet{sepulveda2011} suggest that this large-scale outflow is also excited by one or more of the Class 0 sources.

Using measurements of polarised extinction of background starlight, \citet{sen2005} found a mean magnetic field direction in CB 54 of $116^{\circ}\pm 38^{\circ}$ E of N.  \citet{bertrang2014} found a complex geometry, with near-infrared (NIR) extinction polarisation vectors on the northern side of the globule broadly aligned with the \citet{yun1994} CO outflow axis.
However, \citet{sen2021}, using R-band extinction polarisation, find a random distribution of polarisation angles, suggesting significant amounts of foreground extinction.  \citet{henning2001} observed CB 54 in 850$\mu$m polarised emission with the SCUPOL polarimeter on the James Clerk Maxwell Telescope (JCMT), finding a somewhat random distribution of polarisation vectors.
\citet{wolf2003} re-analysed these data, and found a weak correlation between the magnetic field and large-scale outflow directions in CB 54.

In this paper we present JCMT 850$\mu$m observations of CB 54, made in polarised light using the POL-2 polarimeter.  We describe our observations and the data reduction process in Section~\ref{sec:observations}.  In Section~\ref{sec:results} we discuss the magnetic field geometry, mass and density properties and energetic balance of the globule, and make comparison to previous observations of the globule.  In Section~\ref{sec:discussion} we discuss our results.  Section~\ref{sec:conclusions} summarises this work.

\begin{figure*}
    \centering
    \includegraphics[width=0.8\textwidth]{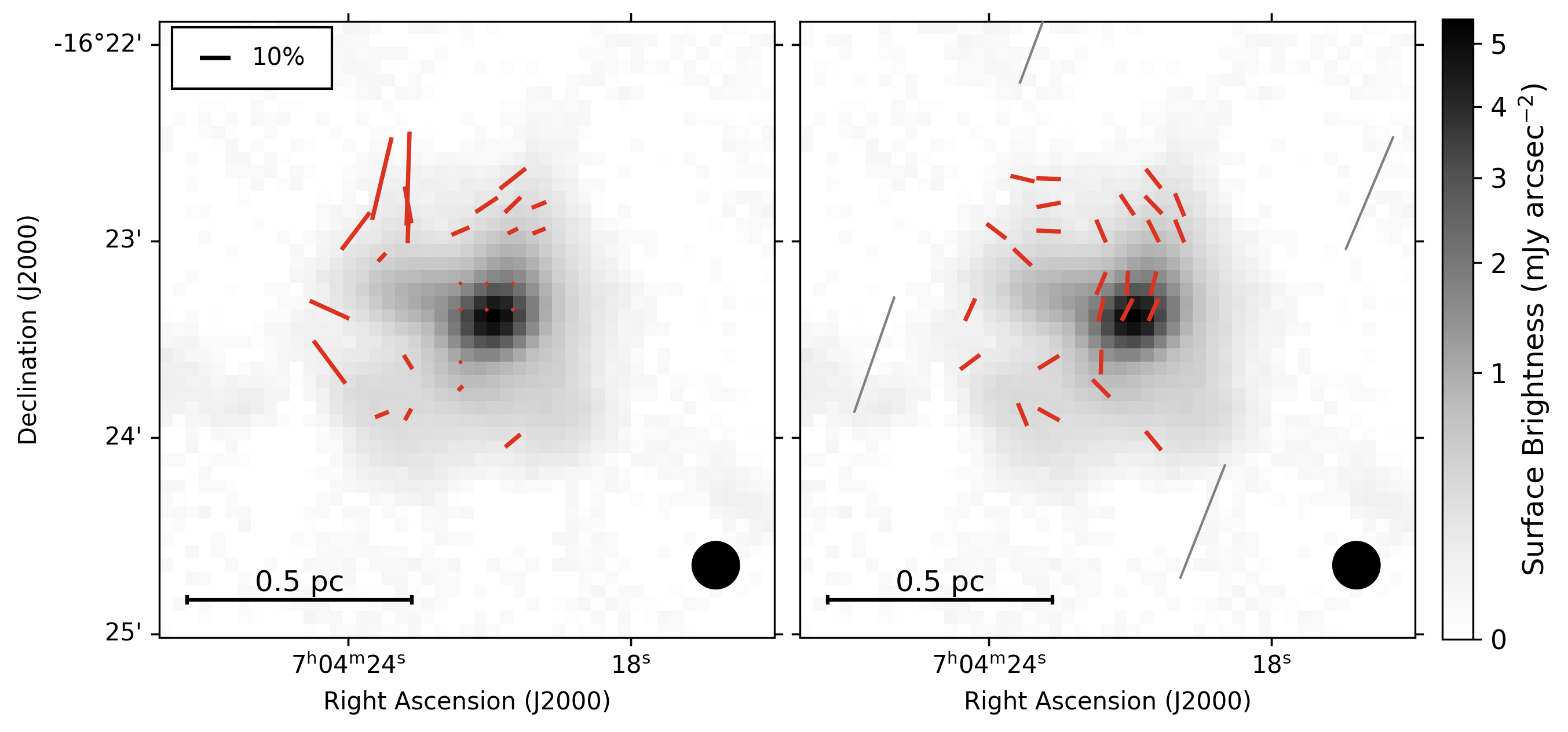}
    \caption{Our observations of CB 54.  Left panel shows polarisation vectors ($I/\delta I > 5$, $p/\delta p > 3$), scaled by polarisation fraction.  Right panel shows uniform-length magnetic field vectors (polarisation vectors rotated by 90$^{\circ}$).  Red vectors are POL-2 measurements; grey vectors are \textit{Planck} measurements which, with an effective resolution of $15^{\prime}$, are significantly oversampled.  The background image is POL-2 850$\mu$m Stokes $I$ emission.  The JCMT beam size is shown in the lower right of each panel.}
    \label{fig:cb54}
\end{figure*}

\section{Observations}
\label{sec:observations}

We observed the Bok globule CB 54 15 times between 2019 March 03 and 2021 February 26 using the POL-2 polarimeter \citep{friberg2016} mounted on the Submillimetre Common-User Bolometer Array 2 (SCUBA-2; \citealt{holland2013}) on the James Clerk Maxwell Telescope (JCMT).  The data were taken in Band 2 weather ($0.05<\tau_{225\,{\rm GHz}}<0.08$) under project codes M19AP016, M20AP022 and M21AP028.  Each observation consisted of a 42-minute POL-2-DAISY scan pattern.

The data were reduced using the $pol2map$\footnote{\url{http://starlink.eao.hawaii.edu/docs/sun258.htx/sun258ss73.html}} script recently added to the \textsc{Smurf} package in the $Starlink$ software suite \citep{chapin2013}.  See \citet{pattle2021} for a detailed description of the current POL-2 data reduction process. Instrumental polarisation (IP) was corrected for using the `August 2019' IP model\footnote{\url{https://www.eaobservatory.org/jcmt/2019/08/new-ip-models-for-pol2-data/}}.  The 850$\mu$m data were calibrated using a flux conversion factor (FCF) of 2795 mJy\,arcsec$^{-2}$\,pW$^{-1}$ using the post-2018 June 30 SCUBA-2 FCF of 2070 mJy\,arcsec$^{-2}$\,pW$^{-1}$ \citep{mairs2021} multiplied by a factor of 1.35 to account for additional losses in POL-2 \citep{friberg2016}.  POL-2 observes simultaneously at 850$\mu$m and 450$\mu$m, but we consider only the 850$\mu$m observations in this work as the 450$\mu$m data reduction process remains under development.

We binned our output vector catalogue to 8-arcsec (approximately Nyquist-sampled) pixels.  The per-pixel RMS noise values in the vector catalogue were then remodelled using the $pol2noise$ script, which models map variance as the sum of three components, based on exposure time, the presence of bright sources, and residuals.  The average RMS noise in Stokes $Q$ and $U$ and $I$ in the central 3 arcmin of the map on 8-arcsec pixels is 0.004 mJy\,arcsec$^{-2}$ (1.0 mJy\,beam$^{-1}$).

The observed polarised intensity is given by
\begin{equation}
    PI^{\prime} = \sqrt{Q^{2} + U^{2}}.
\end{equation}
We debiased this quantity using the modified asymptotic estimator \citep{plaszczynski2014,montier2015}:
\begin{equation}
    PI = PI^{\prime} - \frac{1}{2}\frac{\sigma^{2}}{PI^{\prime}}\left(1-e^{-\left(\frac{PI^{\prime}}{\sigma}\right)^{2}}\right),
\end{equation}
where $\sigma^{2}$ is the weighted mean of the variances {$\sigma_{Q}^{2}$ and $\sigma_{U}^{2}$},
\begin{equation}
    \sigma^{2} = \frac{Q^{2}\sigma_{Q}^{2} + U^{2}\sigma_{U}^{2}}{Q^{2} + U^{2}},
\end{equation}
calculated on a pixel-by-pixel basis. Debiased polarisation fraction is given by $p = PI/I$.

Polarisation angle is given by
\begin{equation}
    \theta_{p} = 0.5\arctan(U,Q).
\end{equation}
We note that the polarisation angles which we detect are not true vectors, as they occupy a range in angle $0-180^{\circ}$.  We nonetheless refer to our measurements as vectors for convenience, in keeping with the general convention in the field.

Throughout this work we assume that dust grains are aligned with their major axis perpendicular to the magnetic field direction \citep[e.g.][]{andersson2015}, and so that the plane-of-sky magnetic field direction can be inferred by rotating $\theta_{p}$ by 90$^{\circ}$.

\section{Results}
\label{sec:results}

850$\mu$m polarisation and magnetic field vector maps are shown in Figure~\ref{fig:cb54}.  In these figures and throughout the following analysis, we use the vector selection criteria $p/\delta p > 3$ and $I/\delta I > 5$.  Note that $p/\delta p > 3$ is equivalent to $\delta\theta_{p}<9^{\circ}.6$ \citep[cf.][]{serkowski1962}.  Maps of Stokes $Q$ and $U$ emission, and of $PI^{\prime}$, are shown in Appendix~\ref{sec:appendix}.

\subsection{Magnetic field geometry}

\begin{figure}
    \centering
    \includegraphics[width=0.47\textwidth]{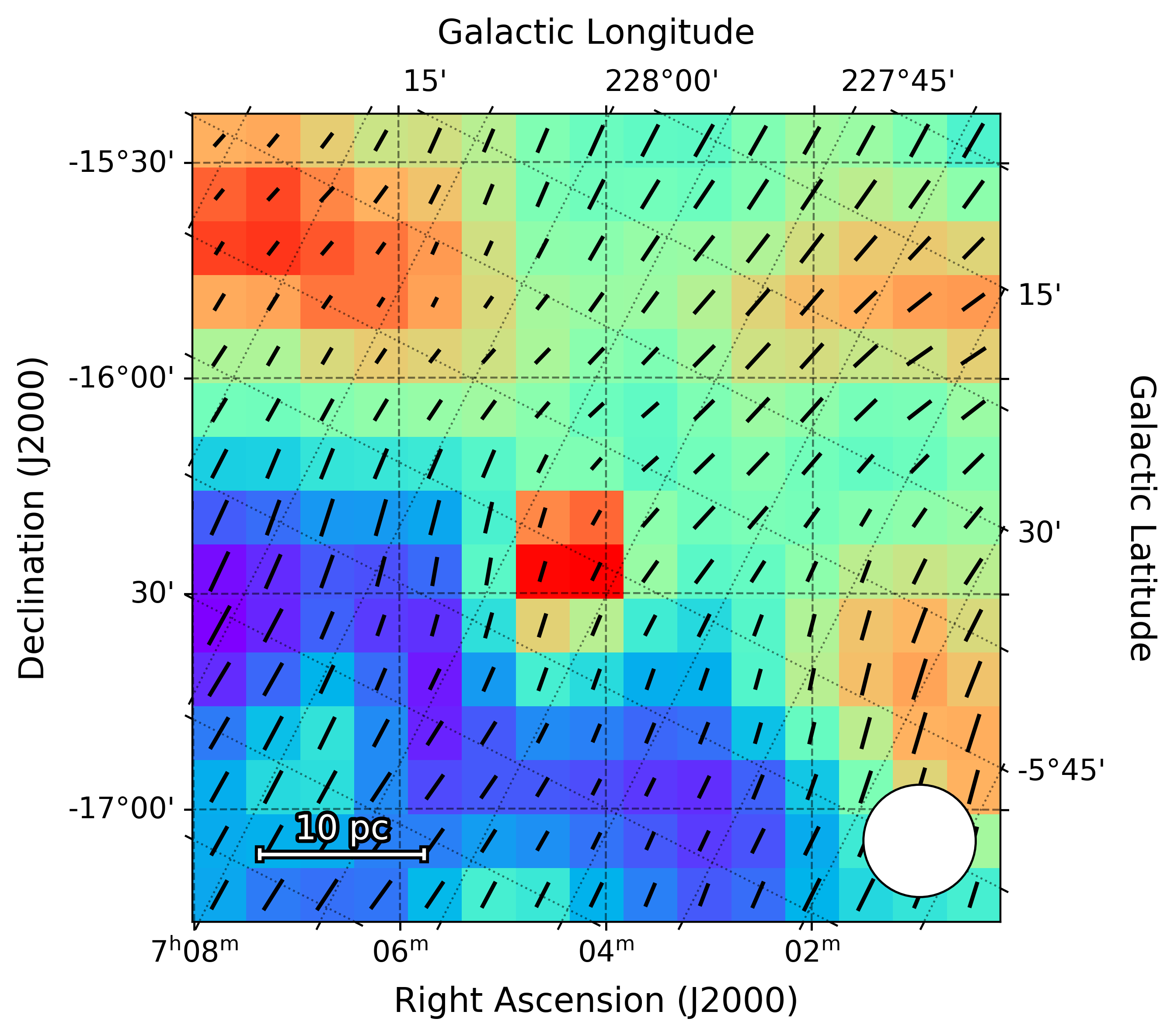}
    \caption{\textit{Planck} magnetic field vectors in CB 54, using the GNLIC foreground dust model \citep{planckGNLIC}.  Vector length is proportional to polarisation fraction.  All \textit{Planck} vectors shown have $p/dp > 3$.  Dashed grid lines show equatorial projection; dotted grid lines show galactic projection.  The effective resolution of this data set is 15$^{\prime}$, and the pixels are Nyquist-sampled. The beam is shown in the lower right-hand corner.}
    \label{fig:planck}
\end{figure}

The 6 vectors which we detect in the dense centre of CB 54 are quite uniform, with a mean position angle $\theta_{B} = 163^{\circ}\pm 8^{\circ}$.  However, in the periphery of the globule, the magnetic field geometry appears curved, showing well-ordered deviation from the field direction which we measure in the centre of the globule.  The field in the periphery of the globule has a mean position angle of $67^{\circ}\pm 45^{\circ}$, and a median position angle of $47^{\circ}$.

We compared our POL-2 data to \textit{Planck} 353\,GHz dust polarisation maps of CB 54 \citep{planckXIX,planckGNLIC}.  The \textit{Planck} archive measures polarisation angle relative to Galactic north, and uses the convention $\theta_{p} = 0.5\arctan(-U,Q)$.  We calculated the difference in polarisation angle between galactic and equatorial systems, $\psi$, using the relation
\begin{equation}
    \psi = \theta_{p, gal}-\theta_{p, eq} = \arctan\left(\frac{\cos(l-32^{\circ}.9)}{\cos b\cot 62^{\circ}.9 - \sin b\sin(l-32^{\circ}.9)}\right)
\end{equation}
\citep{corradi1998}.  For $(l,b) = (228^{\circ}.99,-4^{\circ}.62)$, $\psi = -63^{\circ}.08$.

\textit{Planck} observations of CB 54 are shown in Figure~\ref{fig:planck}.  We used the GNLIC foreground dust model \citep{planckGNLIC}, as it provides covariance matrices for each Stokes parameter, but the Commander dust model \citep{planckXIX} produces near-identical polarisation angles in CB 54.  The mean field direction in the \textit{Planck} data is $158^{\circ}\pm 7^{\circ}$ E of N, consistent with the value of $163^{\circ}\pm 8^{\circ}$ that we measure in the densest part of the globule with POL-2.  Both of these values are also consistent within uncertainties with the mean field direction of $116^{\circ}\pm 38^{\circ}$ measured in the region by \citet{sen2005}.

\begin{figure}
    \centering
    \includegraphics[width=0.47\textwidth]{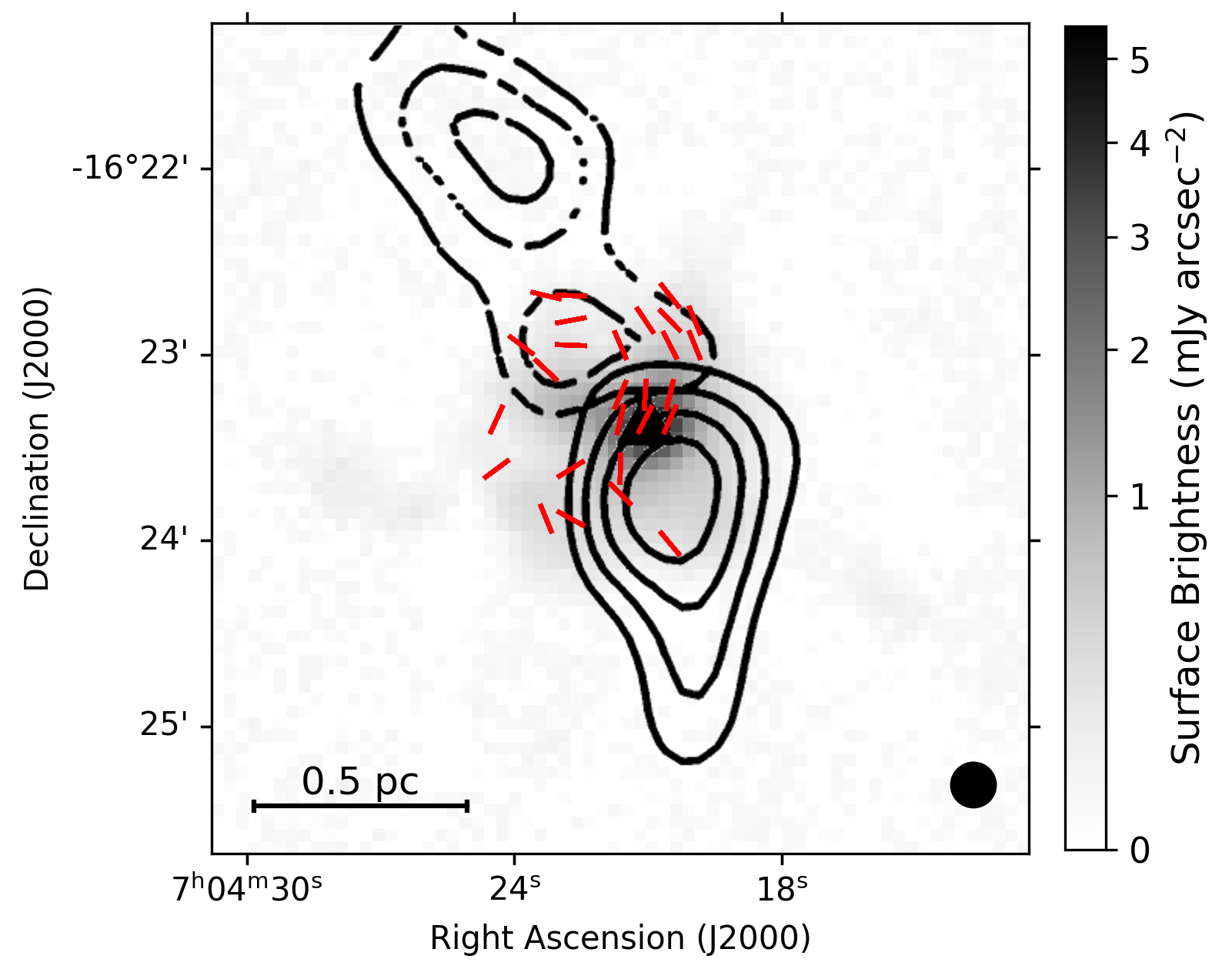}
    \caption{The large-scale $^{12}$CO outflow mapped by \citet{yun1994}, overlaid on our POL-2 data.  Note similarity between magnetic field and outflow morphology. Southern lobe (solid contours) is blue-shifted emission (13--17\,km\,s$^{-1}$), northern lobe (broken contours) is red-shifted emission (23--27\,km\,s$^{-1}$).  Background image shows POL-2 Stokes $I$ emission.}
    \label{fig:big_outflow}
\end{figure}

Although the magnetic field which we observe in the centre of CB 54 is {well-ordered} and consistent with the \textit{Planck}-scale field, the magnetic field direction in the periphery of the globule deviates significantly from that seen in the centre.  There is a striking similarity between the magnetic field morphology in the periphery of CB 54 and the contours of the outflow described by \citet{yun1994}, as shown in Figure~\ref{fig:big_outflow}.  This is most apparent in the northern (red) wing of the outflow, as more polarised emission is detected on the north-eastern side of the globule, and is consistent with previous near-infrared extinction polarisation measurements on larger scales \citep{bertrang2014}.  The few vectors which we detect on the south-western side of the globule are also broadly aligned with the blue wing of the outflow.

On the eastern periphery of the globule, away from the outflow, there are a small number of POL-2 vectors which appear broadly aligned with the \textit{Planck}-scale field.

\subsection{Comparison with SCUPOL}

\begin{figure}
    \centering
    \includegraphics[width=0.47\textwidth]{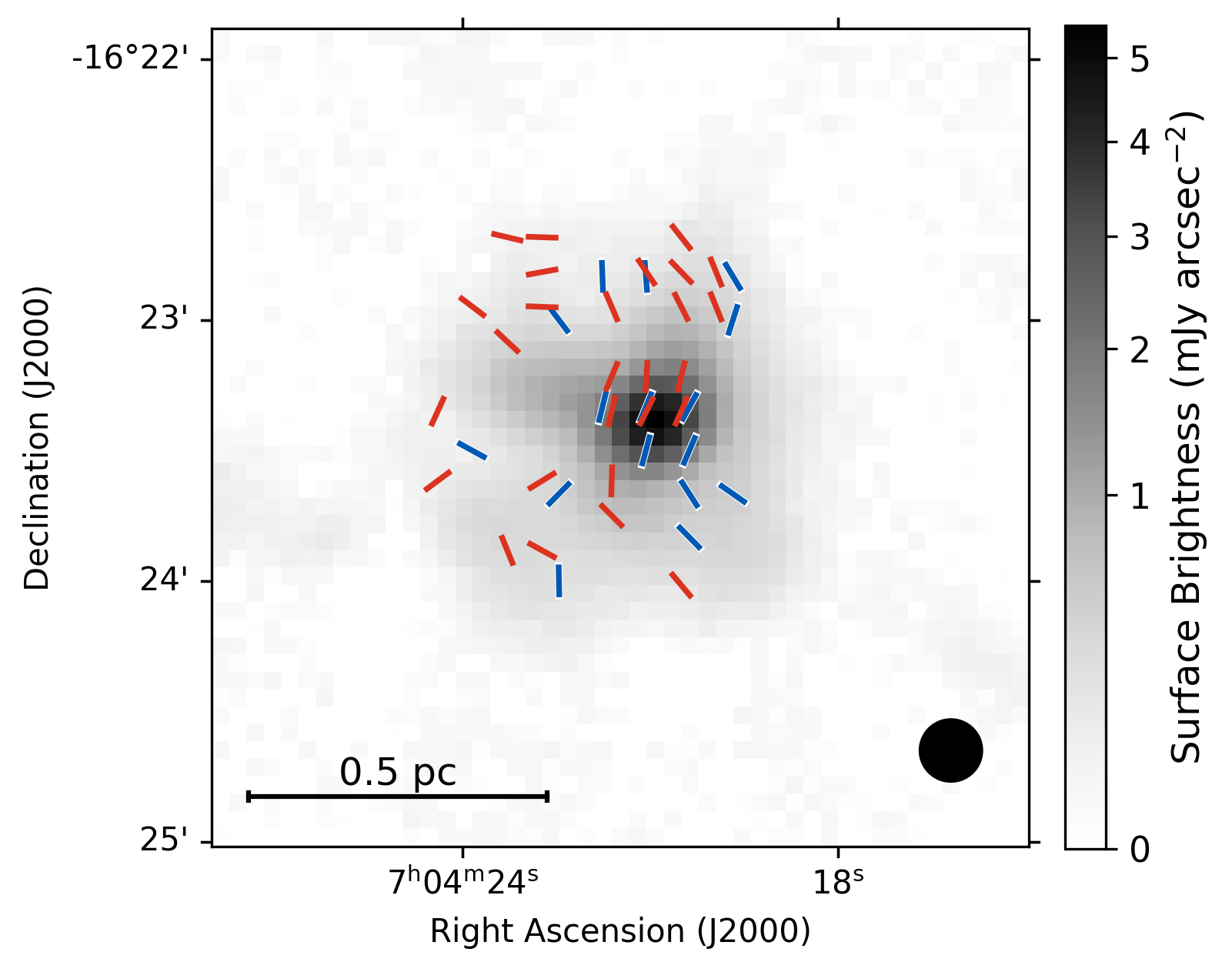}
    \caption{Comparison of POL-2 (red; this work) and SCUPOL (blue with white outline; \citealt{matthews2009}) magnetic field vectors in CB 54.  Background image shows POL-2 Stokes $I$ emission.}
    \label{fig:scupol_pol2}
\end{figure}

\begin{figure}
    \centering
    \includegraphics[width=0.47\textwidth]{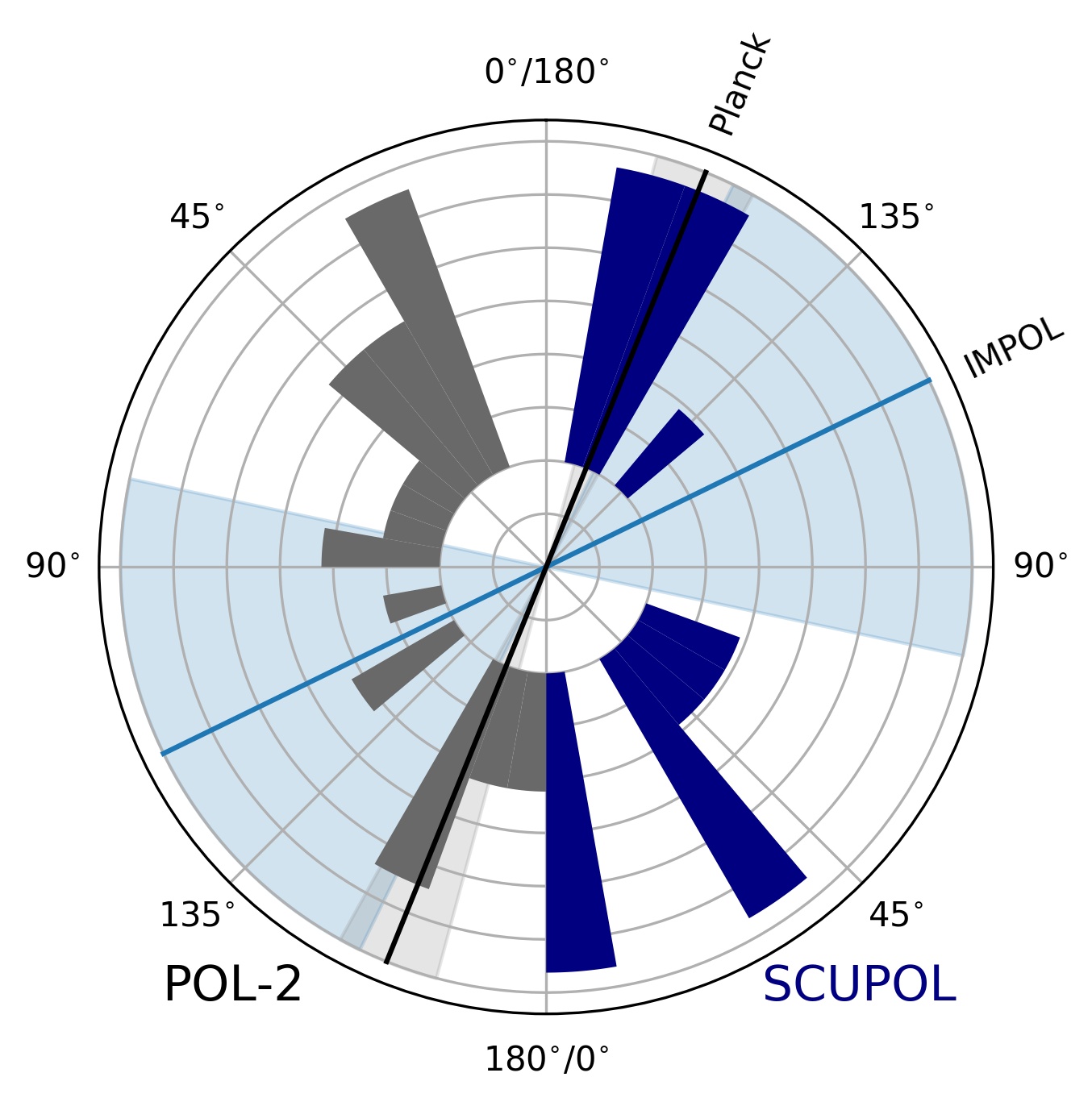}
    \caption{Circular histograms of POL-2 (grey) and SCUPOL (blue) magnetic field angles.  Mean \textit{Planck} and IMPOL \citep{sen2005} magnetic field angles across CB 54 are shown, with their uncertainties shown as shaded sectors.  The POL-2 histogram is shown over the angle range $0-180^{\circ}$ E of N, while the SCUPOL histogram is shown over the range $180-360^{\circ}$.  Due to the $\pm 180^{\circ}$ ambiguity on polarization vector measurements, this range is effectively identical to the range $0-180^{\circ}$, and so opposite angles on the plot agree.  In Cartesian space, the area of each histogram is normalised to 1; the projection of the histograms onto a circle means that their areas are distorted.}
    \label{fig:scupol_pol2_hist}
\end{figure}

We compared our POL-2 data with previous measurements of 850$\mu$m polarisation in CB 54, made using the SCUPOL polarimeter.  These data were originally presented by \citet{henning2001}.  We use the {updated reduction} presented by \citet{matthews2009} in the SCUPOL Legacy Catalogue.  These data are presented on 10$^{\prime\prime}$ pixels.  We selected SCUPOL vectors using the criteria $I/dI > 5$, $p/dp > 3$ and $dp<4$\% (cf. \citealt{matthews2009}).  The SCUPOL and POL-2 vectors are compared in Figure~\ref{fig:scupol_pol2}.  The two sets of polarisation angles are similar: in the centre of CB 54, they agree well, with the mean magnetic field angle over the five central SCUPOL vectors being $159^{\circ}\pm 6^{\circ}$ E of N, compared to the $163^{\circ}\pm 8^{\circ}$ E of N which we measure with POL-2.  In the periphery of the globule, the SCUPOL and POL-2 vectors show qualitative similarity, and the significant deviations from the mean direction in globule centre are apparent in both cases.  A two-sided KS test suggests that the two sets of angles are consistent with having been drawn from the same distribution, with a p-value of 0.37.  Histograms of the POL-2 and SCUPOL angles are shown in Figure~\ref{fig:scupol_pol2_hist}.  The SCUBA/SCUPOL and SCUBA-2/POL-2 systems are fully independent, consisting of separate polarimeters, cameras, observing modes, and data reduction algorithms. 

{\citet{henning2001} inferred a plane-of-sky magnetic field strength of $60^{+11}_{-8}\,\mu$G in the higher-density centre of CB 54 using the Davis-Chandrasekhar-Fermi (DCF; \citealt{davis1951}, \citealt{chandrasekhar1953}) method.  However, to do so they used a measured angle dispersion of $42^{\circ}.7^{+11.1}_{-8.0}$, significantly greater than the limit of $\sim 25^{\circ}$ above which the DCF method has been found to become unreliable \citep{ostriker2001}.  \citet{wolf2003} rederived the magnetic field strength in CB 54 to be $104^{+24}_{-21}\,\mu$G, using the same angular dispersion but a hydrogen number density of $n({\rm H}) = 1.5\times 10^{5}\,$cm$^{-3}$.}

\subsection{Globule mass and density}

We measured the total 850$\mu$m flux density in a 2$^{\prime}$-diameter aperture centred on CB 54 to be $4.19\pm 0.25$\,Jy.  The uncertainty on this value is dominated by the 6\% SCUBA-2 850$\mu$m calibration uncertainty \citep{mairs2021}.

We thus calculated the mass of CB 54 using the \citet{hildebrand1983} relation,
\begin{equation}
    M = \frac{F_{\nu}D^{2}}{\kappa_{\nu}B_{\nu}(T)},
\end{equation}
where $M$ is mass, $F_{nu}$ is flux density at frequency $\nu$, $D$ is distance to the source, $\kappa_{\nu}$ is dust opacity, and $B_{\nu}(T)$ is the \textit{Planck} function at temperature $T$.  Taking $D = 1.5$\,kpc, $\kappa_{\nu}=0.0125$\,cm$^{2}$g$^{-1}$ \citep[e.g.][]{johnstone2017} and $T = 15\,$K \citep{sepulveda2011}, this gives a mass of $117\pm7$\,M$_{\odot}$ for CB 54.  We have here considered only the uncertainty on $F_{\nu}$, and so the formal error bars on our measured mass are quite small.  

This value for the mass of CB 54 can only be considered as an approximation.  $D$, $T$ and $\kappa_{\nu}$ are all subject to significant uncertainties.  We adopt a distance of 1.5 kpc to CB 54 \citep{launhardt1997,ciardi2007}; recent extinction measurements support this distance, but find a potential range of distances to the globule of 1.5--2\,kpc \citep{sen2021}.  $\kappa_{\nu}$ is likely to be accurate to within a factor of 2 \citep{ossenkopf1994,roy2014}.  \citet{sepulveda2011} give $T\lesssim 15$\,K in CB 54, as discussed below, so taking $T=15$\,K gives us a lower limit on the mass traced by POL-2.  

We will not have traced the full extent of the dust emission from CB 54 in these observations: SCUBA-2 is insensitive to large-scale structure due to atmospheric filtering effects \citep{chapin2013}, and POL-2 is more so due to its low scanning speed \citep{friberg2016}.
However, the angular size of CB 54 is $\sim 2^{\prime}$, smaller than the maximum allowed size scale in POL-2 observations of $5^{\prime}$ \citep{friberg2016}, and the source is strongly centrally peaked, so this effect may not significantly reduce the measured mass.

Conversely, we will also have some contamination from the $^{12}$CO $J=3\to 2$ line in our Stokes $I$ map \citep{drabek2012}, which will contribute to the measured 850$\mu$m flux.  This is likely to be most significant in the outer parts of the globule, where the dust column density is low, but the net contribution of $^{12}$CO $J=3\to 2$ to SCUBA-2 850$\mu$m emission in star-forming regions is typically $< 20$\% \citep{drabek2012,pattle2015,coude2016}.

Our estimated mass implies an average density of molecular hydrogen over CB 54 of $n({\rm H}_{2})\sim 5.1\times 10^{3}$\,cm$^{-3}$, taking CB 54 to be a uniform sphere, i.e.
\begin{equation}
    n({\rm H}_{2}) = \frac{3M}{4\pi\mu m_{\textsc{h}}r^{3}},
\end{equation}
where we take $\mu=2.8$ to be the mean particle mass.
However, this is an average over the entire globule.  The centre of the globule, in which the Class 0 protostars are located, is significantly higher-density.
We therefore model CB 54 as a two-layered sphere, with a high-density central core embedded in lower-density surroundings.  We took the inner sphere to have a radius of 16 arcsec, in order to include the high-density central region in which the magnetic field direction is uniform, and in which the protostars are embedded, and the outer sphere to have an inner radius of 16 arcsec and an outer radius of 60 arcsec.

We determined the mass and volume density of the inner and outer regions by initially considering the flux in an annular region covering gas in the outer sphere only, with an inner radius of 16 arcsec and an outer radius of 60 arcsec.  The total flux in the annular region was $2.28\pm 0.14$ Jy.  For the values of $D$, $\kappa_{\nu}$ and $B_{\nu}(T)$ listed above, this is equivalent to a mass of $64\pm 4$ M$_{\odot}$.  The average volume density in the annular region, and so in the `outer sphere' in our model of CB 54, is thus given by
\begin{multline}
        n({\rm H}_{2}) = \frac{M}{\mu m_{\textsc{h}}}\left(\frac{4}{3}\pi a^{2}-2\pi a^{2}\sqrt{r^2 - a^2}\right. \\ - \left.\frac{2}{3}\pi\left(r-\sqrt{r^2 - a^2}\right)^2\left(2r+\sqrt{r^2 - a^2}\right)\right)^{-1},
\end{multline}
where $r = 60$\,arcsec and $a=16$\,arcsec, and so we estimate $n({\rm H}_2) = 3.0\times 10^{3}$\,cm$^{-3}$ in the periphery of the globule.

Conversely, the total flux in the central circular aperture, of radius 16 arcsec, is $1.91\pm 0.11$ Jy, equivalent to a mass of $53\pm 3$ M$_{\odot}$.  Using our previous density estimate for gas in the outer sphere, we estimate that the column of emission contained in the annulus represents the sum of contributions from 6\,M$_{\odot}$ of material in the low-density outer sphere, and from 47\,M$_{\odot}$ of material in the high-density inner sphere.  The inner sphere thus has an average density of $1.0\times 10^{5}$\,cm$^{-3}$.

These values are quite consistent with previous measurements of CB 54.  \citet{sepulveda2011} measured a gas temperature of $T\lesssim 15$\,K, a mass of $\gtrsim 62$\,M$_{\odot}$, and an average density in CB 54 of $\sim 3.3\times10^{3}$\,cm$^{-3}$ from NH$_{3}$ emission {(a suitable dense gas tracer, with a critical density of $\sim 2\times 10^{3}\,$cm$^{-3}$;  e.g. \citealt{juvela2012})}.  The NH$_{3}$ contours shown by \citet{sepulveda2011} correspond well to the 850$\mu$m dust emission which we observe.
Our value for the density in the central region is also comparable to previous measurements.  \citet{henning2001} found a density $n({\rm H})=5\times 10^{4}\,$cm$^{-3}$ within the FWHM contour of a Gaussian profile fitted to CB 54, while \citet{wolf2003} found a density of $1.5\times 10^{5}\,$cm$^{-3}$.

\subsection{Energetics analysis}

{In this section we present a crude energetics analysis to determine an approximate upper limit on the magnetic field strength in the periphery of CB 54, if the magnetic field is indeed being reshaped by the large-scale outflow.}

\citet{yun1994} find that the large-scale CO outflow has a mass in its blue wing of $0.55(D/600\,{\rm pc})^{2}$\,M$_{\odot}$, and in its red wing of $0.2(d/600\,{\rm pc})^{2}$\,M$_{\odot}$.  They further find a total momentum of 4.4\,$(D/600\,{\rm pc})^{2}$M$_{\odot}$\,km\,s$^{-1}$, and an energy of $10^{43}(d/600\,{\rm pc})^{2}$\,erg for the outflow.  The length scale of the outflow is 1.4 arcmin, and its dynamical time is $2.8\times 10^{4}(D/600\,{\rm pc})$\,yr.  Taking the distance to CB 54 to be 1.5\,kpc \citep{ciardi2007,sen2021}, the total outflow mass is {4.7\,M$_{\odot}$, the energy is $6.3\times 10^{43}$\,erg}, and the dynamical time is $7\times 10^{4}$\,yr, while the length scale of the outflow is 0.61\,pc.

From Figure~\ref{fig:big_outflow}, we estimate an aspect ratio in the red lobe of the outflow of $\sim 0.5:1$, and of $\sim 1:1$ in the blue lobe.  If we consider each lobe as a cylinder of height 0.61\,pc, we estimate a total volume of the outflow of $\sim 6.6\times10^{54}$\,cm$^{3}$.  The estimated average energy density in the outflow is thus {$u_{outflow}\sim 1\times 10^{-11}$\,erg\,cm$^{-3}$}.  Alternatively, if we consider each lobe of the outflow as a right circular cone of the same height, we estimate a volume of $\sim 2.2\times 10^{54}$\,cm$^{-3}$, and so an average energy density of {$u_{outflow}\sim 3\times 10^{-11}$\,erg\,cm$^{-3}$}. 

If the magnetic field is reshaped by the outflow, it implies that the magnetic energy density in CB 54 {($u_{B}$)} is less than the outflow energy density, {i.e.}
\begin{equation}
    u_{B} < u_{outflow}.
    \label{eq:ineq}
\end{equation}
Magnetic energy density, $u_{B}$, is given in cgs units by
\begin{equation}
    u_{B} = \frac{B^{2}}{8\pi},
\end{equation}
where $B$ is magnetic field strength, {and so equation~\ref{eq:ineq} is equivalent to placing an upper limit on $B$, such that}
\begin{equation}
    B < \sqrt{8\pi u_{outflow}}.
\end{equation}
{Our estimate of $u_{outflow}$ implies $u_{B} < 1 - 3\times 10^{-11}$\,erg\,cm$^{-3}$}, and so that {$B < 16\,\mu$G (cylindrical outflow geometry) or $B < 27\,\mu$G (conical outflow geometry)} in the low-density peripheral regions of CB 54 where the field appears to have been reshaped by the outflow.  {We therefore adopt an upper limit on the magnetic field strength in the periphery of CB 54 of $B < 27\,\mu$G.}
 
 We note that {while} this {upper limit is very approximate, it is relatively insensitive to the uncertainty on the distance to CB 54.  If we take CB 54 to be at a distance of 1.1\,kpc rather than 1.5\,kpc \citep{brand1993}, the upper limit on $B$ is modified by a factor $\sqrt{1.5/1.1} = 1.16$, becoming $B < 19\,\mu$G (cylindrical outflow) or $B < 32\,\mu$G (conical outflow).  Conversely, taking the upper-limit distance to CB 54 of 2\,kpc \citep{sen2021} leads to $B < 14\,\mu$G or $B < 23\,\mu$G.}

\citet{crutcher2010} used archival Zeeman measurements to derive a maximum magnetic field strength as a function of gas density of
\begin{equation}
    B_{max} = 10\left(\frac{n({\rm H})}{300\,{\rm cm^{-3}}}\right)^{0.65}\,\mu{\rm G}
\end{equation}
for hydrogen number densities $n({\rm H}) > 300$\,cm$^{-3}$.  For our estimated gas density of $n({\rm H}_{2})\sim 3.0\times 10^{3}$\,cm$^{-3}$ in the periphery of CB 54 ($n({\rm H})\sim 6.0\times 10^{3}$\,cm$^{-3}$), this implies $B_{max} = 69\,\mu$G.  Our inferred {upper limit,} $B < 27\,\mu$G, is below {the more general \citet{crutcher2010}} upper limit {at this gas density, and so is consistent with it, and also suggests that the magnetic field in the periphery of CB 54 is comparatively weak}.

\begin{figure*}
    \centering
    \includegraphics[width=0.7\textwidth]{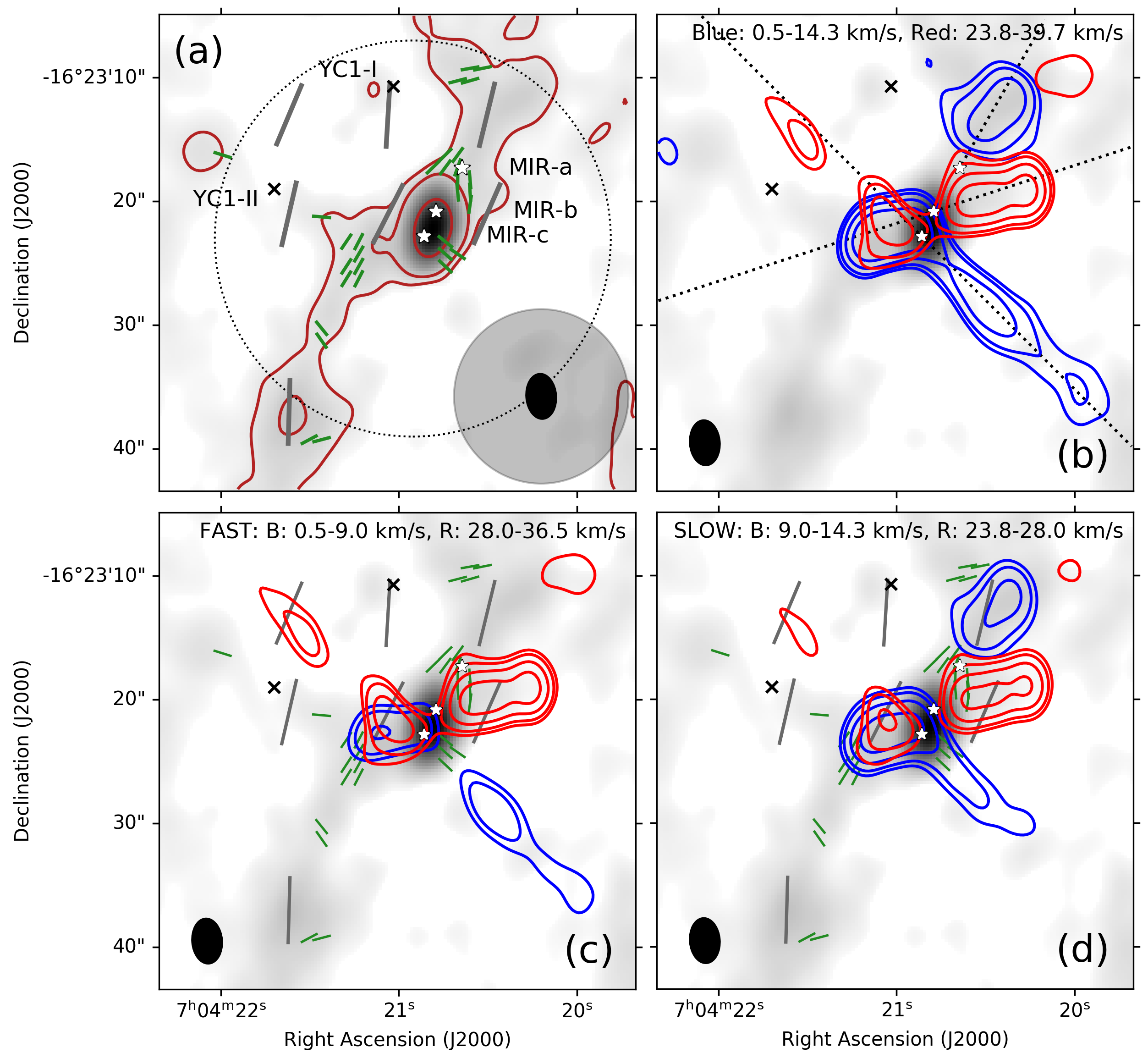}
    \caption{CARMA TADPOL observations of CB 54 \citep{hull2014}.  Background shows 1.3\,mm dust continuum.  Green vectors show magnetic field direction inferred from 1.3\,mm dust polarisation observations.  Grey vectors show POL-2 magnetic field vectors, as presented in Figure~\ref{fig:cb54}.  Blue and red contours in panels (b)-(d) show the $^{12}$CO $J=2-1$ transition, integrated over velocity ranges away from the systemic velocity of CB 54 (as determined by \citealt{hull2014}).  Contour intervals: 2, 3, 5 and 8-$\sigma$, where $\sigma = 1.28$\,mJy/CARMA beam.  (a) CARMA 1.1mm continuum, with embedded sources labelled.  Maroon contours show 3\%, 10\% and 50\% of the maximum brightness, emphasizing the extent and direction of the ridge structure in which the protostars are embedded.  Dotted circle shows the 16$^{\prime\prime}$-radius area over which the central density is estimated in the POL-2 data.  (b) contours of $^{12}$CO emission integrated in the velocity ranges $0.5-14.3\,$km\,s$^{-1}$ (blue) and $23.8-39.7$\,km\,s$^{-1}$ (red) are shown.  The direction of the outflow identified by \citet{hull2014} (108$^{\circ}$ E of N) is shown as a dotted line through MIR-b. Our proposed approximate directions of the outflow from MIR-c, and the blue wing of the outflow from MIR-a are shown as dotted lines through the respective sources.  Magnetic field vectors are excluded for clarity.  (c) contours of `fast' $^{12}$CO emission, integrated in the ranges $0.5-9.0\,$km\,s$^{-1}$ and $28.0-36.5$\,km\,s$^{-1}$ are shown.  (d) contours of `slow' emission, integrated in the ranges $9.0-14.3\,$km\,s$^{-1}$ and $23.8-28.0$\,km\,s$^{-1}$ are shown.  The CARMA beam is shown in each panel.  The JCMT beam is shown in the top left panel.}
    \label{fig:tadpol}
\end{figure*}

\subsection{CARMA observations}

\begin{figure}
    \centering
    \includegraphics[width=0.47\textwidth]{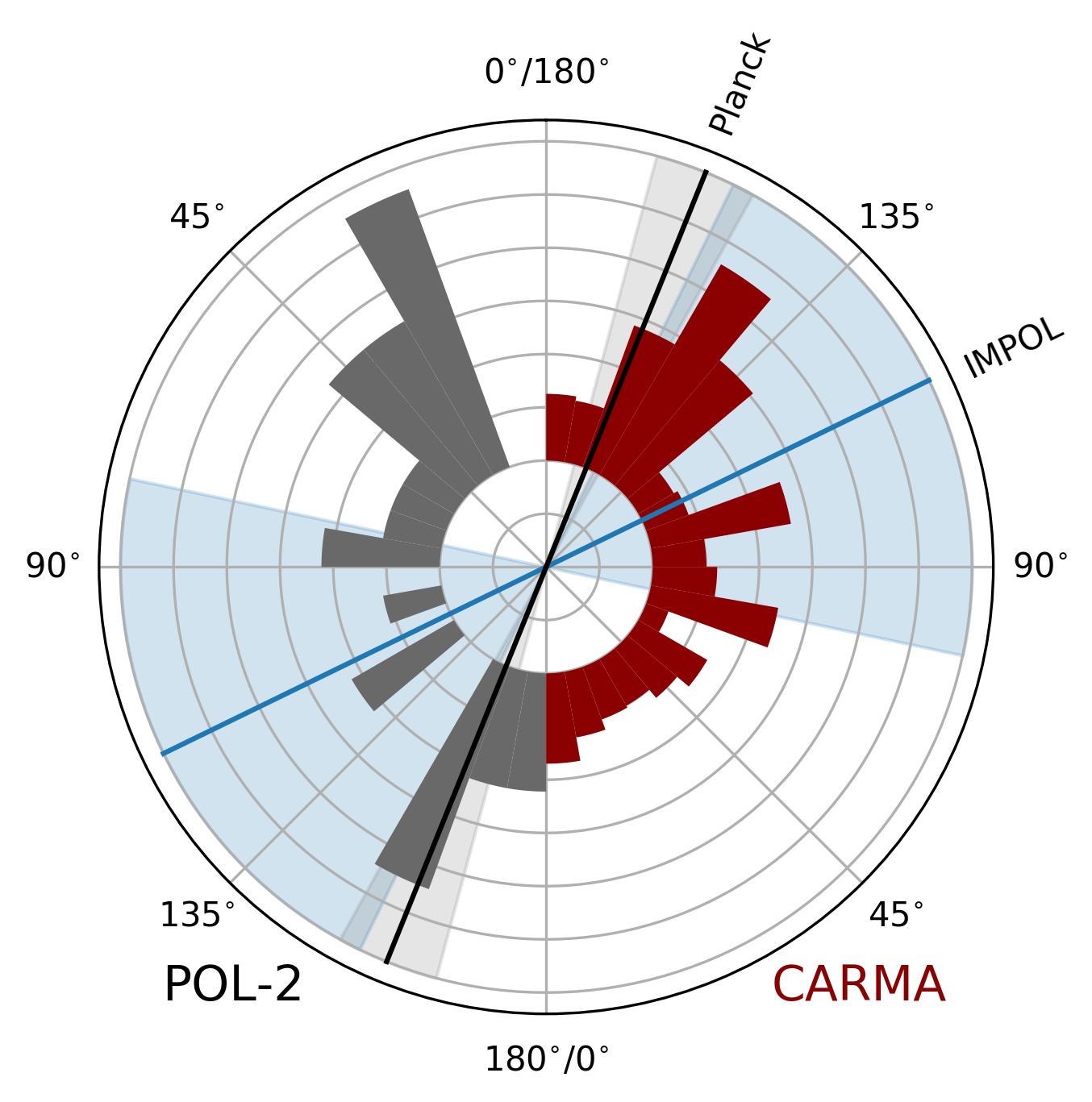}
    \caption{Circular histograms of POL-2 (grey) and CARMA (orange) magnetic field angles.  Opposite angles agree, as described in the caption of Figure~\ref{fig:scupol_pol2_hist}.  Mean \textit{Planck} and IMPOL \citep{sen2005} magnetic field angles across CB 54 are shown, with their uncertainties shown as shaded sectors.}
    \label{fig:pol2_carma_hist}
\end{figure}

The \citet{yun1994} outflow emanates from the dense ridge at the centre of CB 54.  \citet{ciardi2007} detected three Class 0 protostars in this ridge, MIR -a, -b and c, with masses of $\sim 4\,$M$_{\odot}$, $\sim 1.5\,$M$_{\odot}$ and $> 0.2\,$M$_{\odot}$ respectively.  This ridge was mapped in both 1.3 mm dust polarisation and the $^{12}$CO $J=2-1$ transition (230.538\,GHz) at $3^{\prime\prime}.64\times2^{\prime\prime}.41$ resolution by \citet{hull2014} using the CARMA (Combined Array for Research in Millimeter-wave Astronomy) interferometer as part of the TADPOL survey.  We show the \citet{hull2014} CARMA data in Figure~\ref{fig:tadpol}.  The magnetic field in the dense ridge in which the three protostars are forming appears to broadly lie along the length of the ridge (which itself is oriented $\sim 145^{\circ}$ E of N, as shown in Figure~\ref{fig:tadpol}), and is similar to the {magnetic field direction traced by POL-2 on larger scales in CB 54, and with the average field direction in the globule given by \textit{Planck}}.  The magnetic field is approximately perpendicular to the ridge in the material to the northeast and southwest of the ridge, {and parallel to the large-scale outflow direction, again consistent with the behaviour seen on larger scales in the POL-2 data}.  Histograms of POL-2 and CARMA magnetic field angles are shown in Figure~\ref{fig:pol2_carma_hist}.

Inspection of the CARMA $^{12}$CO data presented by \citet{hull2014} suggests that there is (i) a fast, highly-collimated but quite faint outflow running $\sim 45^{\circ}$ E of N, which appears most likely to be excited by MIR-c; (ii) a slower, broader, but brighter outflow running 108$^{\circ}$ E of N (identified by \citealt{hull2014}), which appears to be excited by MIR-b; (iii) velocity structure suggestive of an outflow around MIR-a, the red wing of which is either not visible or not distinguishable from the MIR-b outflow, but the blue wing of which is oriented $\sim 135^{\circ}$ E of N.

It appears from this that the large-scale CO outflow in CB 54 \citep{yun1994} is associated with the fast, collimated outflow which appears likely to arise from MIR-c.  MIR-c is the faintest of the three Class 0 sources in the MIR, with a lower-limit mass of $> 0.2$\,M$_{\odot}$ \citep{ciardi2007}.  This suggests that MIR-c is young and deeply embedded in the dense centre of CB 54.

The total mass in the large-scale outflow is $\sim 4.7\,$M$_{\odot}$ \citep{yun1994}, significantly greater than the $0.2$\,M$_{\odot}$ lower limit placed on the mass of MIR-c itself by \citet{ciardi2007}.  \citet{machida2012} suggest that in a single protostellar cloud core, 26--40\% of the initial cloud mass is incorporated into the protostar, while 8--49\% is ejected by the outflow.  An outflow mass of 4.7\,M$_{\odot}$ would thus imply a mass reservoir of 9.6--58.8\,M$_{\odot}$ for MIR-c, and so an eventual mass of MIR-c of 2.5--23.5\,M$_{\odot}$.  As there is no indication that CB 54 is a high-mass star-forming region, it seems likely that MIR-c is at the lower end of this potential mass range.  As discussed above, the total mass of CB 54 appears to be broadly in the range $\sim 60-120$\,M$_{\odot}$, sufficiently large to accommodate even the largest value of the implied mass reservoir for MIR-c, but is shared by the three forming protostars.

If MIR-c is indeed a young and deeply embedded Class 0 source, this suggests that the reshaping of the magnetic field by the CO outflow has been rapid, as does the dynamical time of the outflow, $7\times 10^{4}$\,yr \citep{yun1994}.  The magnetic field direction implied by CARMA observations to the northeast and southwest of the dense ridge has a direction similar to that of the outflow emanating into these regions from MIR-c, and perpendicular to the average magnetic field direction in the dense ridge implied by both the POL-2 and CARMA observations.  This supports the suggestion that this outflow is reshaping the magnetic field in the lower-density material of the globule.  Conversely, the magnetic field in the dense material of the ridge does not appear to be reordered by the MIR-c outflow.

\section{Discussion}
\label{sec:discussion}

If the outflow is indeed reshaping the magnetic field on large scales in the globule, this implies a relatively weak magnetic field in CB 54, {significantly less than the likely maximum field strength of $\sim 70\,\mu$G at this gas density implied by the  \citet{crutcher2010} relation}.  We here discuss the extent to which our observations support the suggestion that the magnetic field in CB 54 is {dynamically subdominant.  It is important to note that the relative dynamic importance of magnetic fields with respect to gravity, non-thermal motions and thermal pressure may differ.  These energy balances are typically quantified using the mass-to-flux ratio, Alfv\'en Mach number, and plasma beta respectively \citep[e.g.][and refs. therein]{pattle2022}.  The fact that stars have formed within CB 54 indicates that on at least some size scales, gravity must win out over the other forces involved in the evolution of the globule; however, simulations indicate that the energy balance on larger scales can make both a qualitative and quantitative difference to the evolution of star-forming regions \citep[e.g.][]{seifried2020}.}

\subsection{Density structure}

The magnetic field geometry in CB 54 is {well-ordered} and quite consistent in direction over many orders of magnitude in size scale, except in the vicinity of the \citet{yun1994}/MIR-c outflow.

{Strong-field star formation theory predicts that the minor axis of a dense core will lie parallel to the magnetic field direction \citep{mestel1966,basu2009}.  \citet{basu2000} show that a magnetic field parallel to the minor axis of a triaxial core can have a wide range of orientations in projection, and is most likely to appear at an angle $\sim$10--30$^{\circ}$ to the projected minor axis of the core.  We fitted a two-dimensional Gaussian model to the POL-2 I map of CB54, finding the best-fit model to have a low aspect ratio of 1.15 (eccentricity $e = 0.49$), and a major axis oriented 118$^{\circ}$ E of N.  As discussed above, the mean magnetic field direction which we measure in the centre of CB 54 is $163^{\circ}\pm 8^{\circ}$.  Thus, we infer that the magnetic field observed by POL-2 is offset from the minor axis of the globule by $\sim 45^{\circ}$, in projection.}

{However, performing the same fitting process on the CARMA data shown in Figure~\ref{fig:tadpol} results in a best-fit model with an aspect ratio of 1.56 ($e = 0.77$), and a major axis oriented 164$^{\circ}$ E of N.  The average magnetic field direction in the globule is thus very well-aligned with the major axis of the dense core in which the MIR-a, -b and -c protostars are forming.  While this does not rule out the field being parallel to the minor axis in three dimensions, the probability of such a configuration is extremely low \citep{basu2000}.  The similarity between the magnetic field angle measured by POL-2 and the major axis of the dense core, along with the (sub)millimetre brightness of this core, also suggests that the polarized emission observed by POL-2 may be dominated by unresolved emission from this dense core.}

Many recent observations have shown that dense filaments preferentially align perpendicular to the direction of the local magnetic field, while lower-density filaments tend to align parallel to the magnetic field \citep[e.g.][]{alina2019,soler2019,doi2020,arzoumanian2021,kwon2022}.  Simulations suggest that {a transition from parallel to perpendicular alignment at high gas densities will typically only occur in clouds in which the magnetic field's energetic importance is initially similar to or greater than that of gravity} \citep[e.g.][]{soler2013,seifried2020}.  {\citet{seifried2020} found that for their models with initial magnetic field strengths below $\sim 5\,\mu$G, the magnetic field remained parallel or randomly oriented with respect to both density and column density structures at all gas densities.  However, there is no clear dependence of the transition density from parallel to perpendicular field/filament orientations on Alfv\'en Mach number \citep[][and refs. therein]{pattle2022}.}

Whether the {extended structure seen by CARMA in which the dense star-forming core at the centre of CB 54 is embedded} can be considered as a `filament' in the sense used in nearby molecular clouds (e.g. \citealt{andre2010}) is arguable.  The young protostars in CB 54 are nonetheless {forming within a core which is} embedded in {a structure} which is elongated parallel rather than perpendicular to the local plane-of-sky magnetic field direction. 
{This} would suggest a sub-dominant role for the magnetic field {in comparison to gravity not only in the dense, star-forming core, but also in the evolution of the globule as a whole} \citep[e.g.][]{soler2013}.  It is however possible that the filament and the magnetic field could be perpendicular to one another in three dimensions despite being parallel in projection {(e.g. \citealt{doi2020})}.

{We note that an} alternative explanation of the observed correlation between the outflow and magnetic field directions in the periphery of CB 54 is that the field geometry existed prior to the presence of the outflow, and that this field has shaped and channeled the outflow as it leaves the centre of the globule.  {However, this} would require the field to {have changed direction in the periphery of the globule when it is otherwise consistent in direction over many orders in magnitude in size scale}.  This scenario is difficult to physically motivate, but cannot be ruled out.

We also note that this apparent consistency of the magnetic field direction over several orders of magnitude in size scale suggests a highly ordered magnetic field (away from the outflow).  The presence of such an ordered magnetic field suggests that non-thermal motions in the region are sub-Alfv\'enic \citep[e.g.][]{ostriker2001}.  This implies a relatively strong magnetic field in CB 54 {in comparison} to turbulent motions.  {The magnetic field geometry which we observe in CB 54 thus allows us to hypothesise that in this region, while the dynamics are dominated by gravitational collapse, the magnetic energy in the region is greater than that of the non-thermal gas motions.}

\subsection{Star formation efficiency}

{{Both dynamically important magnetic fields and turbulence are expected to decrease star formation efficiency (SFE) by providing support on large scales in molecular clouds \citep[e.g.][]{price2009,federrath2013}.}  If the magnetic field in CB 54 is indeed not sufficiently strong to support the globule against gravitational collapse, and turbulence in the globule is sub-Alfv\'enic, this suggests that gravitational collapse, and so star formation, in the globule ought to proceed relatively efficiently.  The SFE of a molecular cloud varies from $< 1$\% in regions of distributed star formation to up to $\sim 40$\% in regions forming large stellar clusters, with a mean value of $\sim 15$\% \citep{bonnell2011}.  \citet{ciardi2007} estimate a total mass of 10--15\,M$_{\odot}$ for {all of the sources embedded within} CB 54, although as their estimated mass of MIR-c is a lower limit, so too is this total.  Comparing this to our measured dust mass of $117\pm 7\,$M$_{\odot}$ suggests a current SFE of $M_{\rm embedded}/(M_{\rm embedded} + M_{\rm globule})\gtrsim 8-11$\%.  {As discussed above, the total mass of CB 54 is quite uncertain, but its centrally-condensed geometry suggests that the mass which we estimate is representative of the total mass of the globule.}  The final SFE of CB 54 will be higher {than this estimate}, as star formation is still ongoing and the Class 0 protostars may yet accrete a significant fraction of their final mass from the globule \citep{andre1993}.}

{The current SFE of CB 54 appears to be lower than the mean efficiency of star-forming clouds \citep{bonnell2011}.  However, as discussed by \citet{bonnell2011}, comparison to this mean value can elide the significant differences between expected SFEs in high-mass and low-mass star-forming environments. \citet{yun1990} inferred a typical star formation efficiency of $\sim$6\% in Bok globules.  Star formation in CB 54 thus appears to be proceeding somewhat more efficiently than in the average Bok globule.}

{{More energetic outflows, and more efficient coupling between outflows and magnetic fields, are likely to result in lower SFE in molecular clouds as a whole \citep[e.g.][]{krumholz2019}.}  A total mass in the large-scale outflow {from CB 54} of $\sim 4.7\,$M$_{\odot}$ and a dynamical time of the outflow of $\sim 7\times 10^{4}$\,yr \citep{yun1994} imply a very approximate mass outflow rate of $\sim 7\times 10^{-5}$\,M$_{\odot}$\,yr$^{-1}$.  This is {consistent with the typical outflow rate of $\sim 10^{-4}\,$M$_{\odot}$\,yr$^{-1}$ for young Class 0 sources \citep{bontemps1996}, and} with the expected $\gtrsim 10^{-5}$\,M$_{\odot}$\,yr$^{-1}$ mass accretion rate for Class 0 protostars \citep{whitworth2001}, assuming that the mass accretion and outflow rates are comparable \citep[e.g.][]{machida2012}.  
{This suggests that the large-scale CB 54 outflow is quite typical for an outflow being driven by a Class 0 source such as MIR-c.}  However, {as the relationship between magnetic field strength and outflow rate is non-monotonic \citep{tomisaka2002}, and} given its highly approximate nature, {the outflow rate which we estimate} should not be over-interpreted.}

\subsection{Outflow orientations}

Each of the three outflows in the centre of CB 54 has quite different orientation, as shown in Figure~\ref{fig:tadpol}.  Random outflow alignments, with respect to both one another and to density structure, have been seen in both the Orion \citep{davis2009} and Perseus \citep{lee2016,stephens2017} molecular clouds.  However, \citet{stephens2017} find that the distribution of outflow orientations in Perseus is also consistent with a mixture of parallel and perpendicular alignments with respect to density structure, with perpendicular alignment a factor of 3 more common than parallel.  Conversely, \citet{kong2019} find that in the IRDC G28.37+0.07 there is widespread alignment between outflow directions, with outflows having a strong preference to being perpendicular to the IRDC axis.  A dynamically important magnetic field is likely to be a necessary condition for outflows to have a consistent orientation \citep{li2018}.  However, random outflow directions could be created by interactions within multiple systems despite the presence of a strong magnetic field \citep{offner2016}, and so the lack of {a consistent outflow direction amongst the protostars} in CB 54 is not necessarily indicative of a weak magnetic field {in the region}.

{Ideal magnetohydrodynamic (MHD) simulations have shown that magnetic braking is more efficient in protostellar systems whose rotation axis is perpendicular to the magnetic field direction than in those whose rotation axis and magnetic field directions are parallel \citep{matsumoto2004,hennebelle2009}.  This efficient magnetic braking thus favours formation of a larger disc, and a weaker outflow, as misalignment between the field and rotation axes increases \citep{matsumoto2004}.  Creating a significant outflow in the highly misaligned case requires the system to have a high ratio of thermal kinetic to gravitational potential energy \citep{tsukamoto2018}.  Non-ideal MHD simulations can produce an outflow even in the fully misaligned case \citep{hirano2020}, albeit a weak one.}

{The large-scale outflow from CB 54 appears to be significantly misaligned with the magnetic field direction in the core.  While we cannot determine the three-dimensional angle between the field and the outflow, the offset which we see in projection is near 90$^{\circ}$.  This extended outflow apparently perpendicular to the magnetic field direction appears to be unusual in the light of the predictions of both ideal and non-ideal simulations \citep{tsukamoto2018,hirano2020}, and so makes MIR-c a strong candidate for higher-resolution interferometric follow-up, in order to determine whether its magnetic field direction on protostellar core scales is significantly offset from that in its surroundings.}

\subsection{Comparison with other regions}

Although we have not demonstrated a causal link between the {outflow direction and the magnetic field geometry in the periphery of CB54}, we note that if the outflow is indeed reshaping the {magnetic field}, this would be consistent both with previous observations of CB 54 \citep{bertrang2014} and with recent results from other regions, {as well as with the predictions for the evolution of magnetic fields in the presence of outflows from numerical modelling \citep{hirano2020}}.

Magnetic fields have been seen to be reshaped by protostellar outflows on scales $\sim 1000$\,au ($\sim 0.005$\,pc) in recent ALMA observations of the BHR 71 IRS 2 source  \citep{hull2020}.  In this case, the magnetic field traced by ALMA polarised dust emission appears to be that of the outflow cavity wall.  However, the neighbouring IRS 1 source, which forms a wide binary with IRS 2, shows an apparently undistorted hourglass-shaped field.  BHR 71 is $\sim 40$\,M$_{\odot}$ Bok globule at a distance of $\sim 200$ pc, with a large-scale collimated outflow on size scales $\sim 0.3$\,pc \citep{bourke1997}.   Although lower-mass, BHR 71 appears to be somewhat analogous to CB 54.  However, \citet{kandori2020} mapped the magnetic field in BHR 71 using SIRPOL NIR polarimetry at the IRSF, finding the field to be significantly distorted from the mean plane-of-sky field direction in the region.  They ascribe this distortion to an interaction between the globule and a large-scale shock \citep[cf.][]{inoue2013}.

\citet{davidson2011} observed the magnetic field structure on $0.04-0.09$\,pc scales around the nearby low-mass B335, L1527, IC348-SMM2 protostars at 350\,$\mu$m using the SHARP polarimeter on the CSO.  They found that the magnetic field geometries observed around L1527 and IC348-SMM2 could be consistent with magnetically-regulated star formation models distorted by bipolar outflows, but that the magnetic field geometry around B335 was not consistent with magnetically-regulated models, suggesting that a significant amount of distortion had occurred.  They gave an outflow energy density of $\sim 10^{-9}\,$erg\,cm$^{-3}$ for B335, and further found that the energy densities of all of the outflows were sufficiently large to distort magnetic fields of strengths predicted by magnetically-regulated star formation models.  They suggested that the relative lack of distortion in L1527 and IC348-SMM2 is due to the youth of these outflows, $\sim 10^{3}$\,yr, compared to $\sim 2\times 10^{4}$\,yr in B335.  POL-2 and ALMA polarisation observations of B335 support the presence of a field aligned with the outflow direction \citep{yen2019}. The \citet{davidson2011} sources are lower in mass than is CB 54, which houses multiple protostars.  However, in both CB 54 and B335 a single outflow from a low-mass protostar appears to be shaping the magnetic field, suggesting an at least qualitative similarity between these sources.

In Orion B NGC 2071 \citep{lyo2021}, the magnetic field appears to be shaped by the outflow from NGC 2071 IRS 3 on $\sim 0.25$\,pc scales, similar to CB 54.  The NGC 2071 IRS 3 outflow, as estimated from $^{13}$CO emission, is significantly more energetic than the CB 54 outflow, with $u_{outflow} = 2.33\pm 0.32\times 10^{-8}$\,erg\,cm$^{-3}$ ($\gtrsim 10^{3}\,\times$ that of CB 54, as discussed above).  \citet{lyo2021} estimate a magnetic field strength of $563\pm421\,\mu$G in the central 0.12\,pc of NGC 2071, and a magnetic energy density $\sim 2\times 10^{-8}$ \,erg\,cm$^{-3}$, comparable to the outflow energy density in the region.

An extreme example of this behaviour is the explosive BN/KL outflow in the OMC-1 region of Orion A, which appears to have substantially reordered the magnetic field in the region on $\sim 0.01$\,pc scales (\citealt{cortes2021}; see also \citealt{pattle2021a}).  However, the BN/KL outflow is an exceptional event, with an energy density of $\sim 6\times 10^{-4}$\,erg\,cm$^{-3}$ over the area over which the magnetic field is reordered \citep{bally2017,bally2020}, more than three orders of magnitude greater than the magnetic energy density in the region \citep{cortes2021}.  Moreover, it is likely to have arisen from the interaction or decay of a multiple system, and so is not representative of usual conditions in star-forming cores.

The magnetic field geometry in dense star-forming cores is a key discriminant between weak- and strong-field modes of star formation \citep[e.g.][]{myers2021}.
Our observations of CB 54 suggest that even a weak outflow may be able to significantly alter the geometry of a weak magnetic field on scales $\gtrsim 0.1\,$pc in the cloud from which its driving protostar formed, and can do so on relatively short timescales, $\sim 10^{4}$\,yr.  This suggests that if we wish to understand the role of magnetic fields in the evolution of molecular clouds to gravitational instability, we must look at regions where stars have not yet formed.  The magnetic field observed in the presence of an outflow-driving protostar cannot be relied upon to inform us of the field in the gas from which the protostar formed.

\section{Conclusions}
\label{sec:conclusions}

We have observed the large Bok globule CB 54 with the POL-2 polarimeter on the SCUBA-2 camera on the James Clerk Maxwell Telescope (JCMT).

We find that the magnetic field in the centre of the globule is oriented $163^{\circ}\pm 8^{\circ}$ E of N.  The field in the periphery of the globule shows significant, ordered deviation from the mean field direction in the globule centre, with a mean orientation of $63^{\circ}\pm 45^{\circ}$ and a median orientation of $47^{\circ}$.  This deviation corresponds qualitatively with the contours of the extended but relatively weak $^{12}$CO outflow identified in the region by \citet{yun1994}, which appears to emanate from the Class 0 source MIR-c at the centre of the globule.

If the magnetic field in the periphery of the globule is indeed reshaped by the outflow, energetics analysis suggests {that $B < 27\,\mu$G in the area}.  This is less than the maximum likely value at this gas density, $B_{max}\approx 69\,\mu$G, given by the \citet{crutcher2010} relation.

Comparison with archival \textit{Planck} and CARMA measurements shows that the field in the centre of the globule is consistent over several orders of magnitude in size scale, but oriented parallel to the density structure in the region in projection.  This leads us to hypothesise that non-thermal motions in the region are sub-Alfv\'enic, but that the magnetic field itself is subdominant to gravity, {suggesting that the gravitational collapse of CB 54 is relatively unimpeded}.  {We estimate a star formation efficiency in CB 54 of $\gtrsim 8-11$\%.  This is below the average value for molecular clouds, but may be relatively efficient for a Bok globule.}

{The outflow from CB 54 is extended despite significant misalignment between the outflow direction and the average magnetic field direction in the globule.  This suggests that the likely driving source, the Class 0 protostar MIR-c, may be unusual, and merits high-resolution interferometric follow-up, to determine the magnetic field geometry in the protostellar envelope.}

Our results suggest that even a weak outflow can significantly reshape weak magnetic fields in star-forming regions on scales $>\,0.1$\,pc, and so that the magnetic field observed in a star-forming core in the presence of outflows cannot be relied upon to inform us of the pre-star-formation magnetic conditions in the region.

\section*{Acknowledgements}

The authors would like to thank Jo\~ao Yun for helpful discussions of his observations of CB 54.  K.P. is a Royal Society University Research Fellow supported by grant number URF\textbackslash R1\textbackslash211322, and in the early stages of this project was supported by the Ministry of
Science and Technology of Taiwan under grant No. 106-2119-M-007-021-MY3. S.P.L. acknowledges grants from the Ministry of
Science and Technology of Taiwan 106-2119-M-007-021-MY3 and 109-2112-M-007-010-MY3. S.C. acknowledges the SOFIA Science Center, operated by the Universities Space Research Association under contract NNA17BF53C with the National Aeronautics and Space Administration. S.W. and N.Z. acknowledge the support by the DLR/BMBF grant 50OR1910. W.K. was supported by the National Research Foundation of Korea (NRF) grant funded by the Korea government (MSIT) (NRF-2021R1F1A1061794).  C.W.L. is supported by the Basic Science Research Program through the NRF funded by the Ministry of Education, Science and Technology (NRF- 2019R1A2C1010851).  The James Clerk Maxwell Telescope is operated by the East Asian Observatory on behalf of The National Astronomical Observatory of Japan; Academia Sinica Institute of Astronomy and Astrophysics; the Korea Astronomy and Space Science Institute; the National Astronomical Research Institute of Thailand; Center for Astronomical Mega-Science (as well as the National Key R\&D Program of China with No. 2017YFA0402700). Additional funding support is provided by the Science and Technology Facilities Council of the United Kingdom and participating universities and organizations in the United Kingdom, Canada and Ireland.  Additional funds for the construction of SCUBA-2 were provided by the Canada Foundation for Innovation. The authors wish to recognize and acknowledge the very significant cultural role and reverence that the summit of Maunakea has always had within the indigenous Hawaiian community.  We are most fortunate to have the opportunity to conduct observations from this mountain.

\section*{Data Availability}

The raw data used in this analysis are available in the JCMT archive at the Canadian Astronomy Data Centre (CADC) under project codes M19AP016, M20AP022 and M21AP028.  The reduced data presented in this paper are available from the CADC at \url{https://doi.org/10.11570/22.0020}.



\bibliographystyle{mnras}




\appendix
\section{Stokes maps}
\label{sec:appendix}

\begin{figure}
    \centering
    \includegraphics[width=0.47\textwidth]{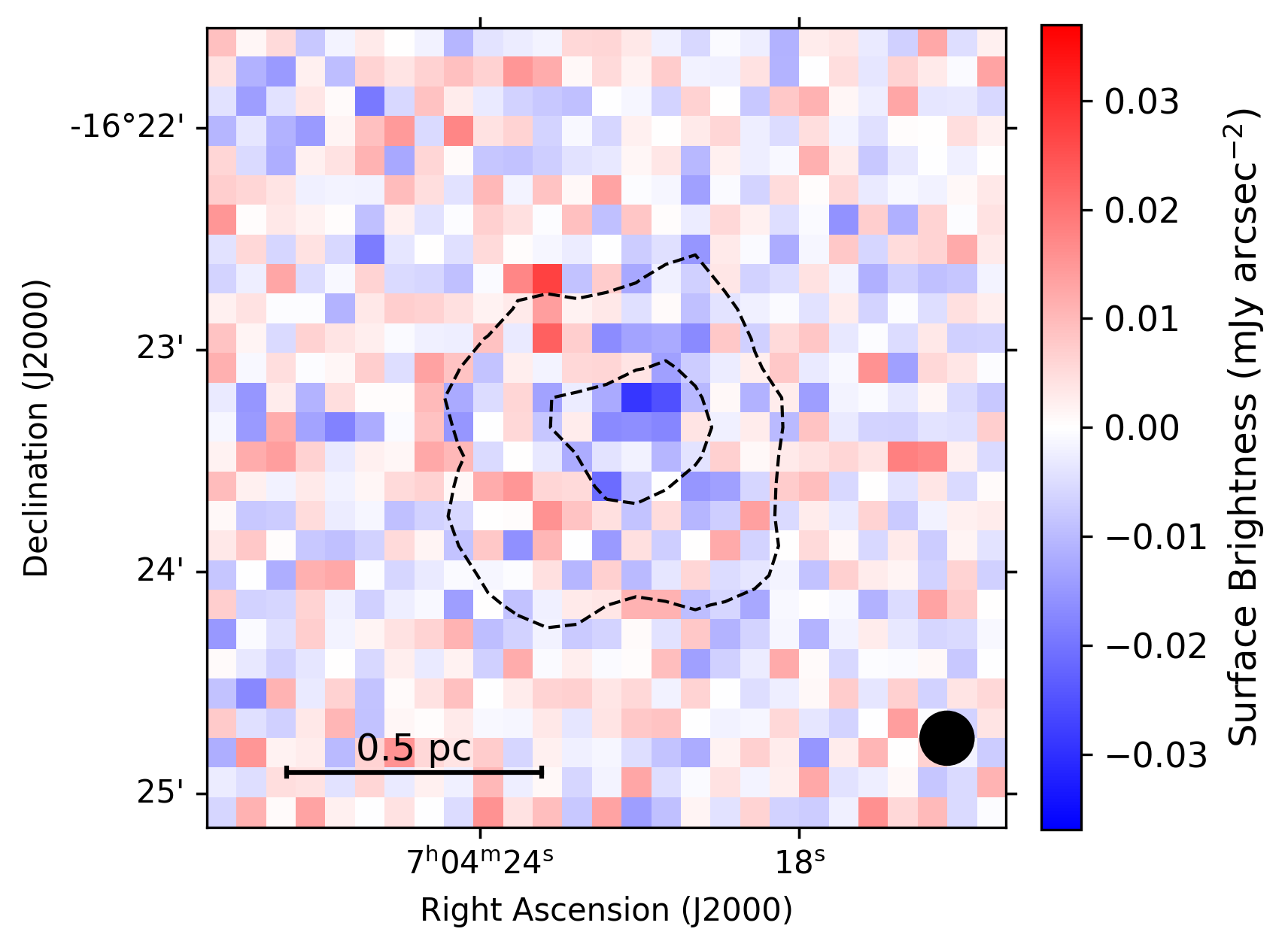}
    \caption{Stokes Q emission in CB 54, shown on 8$^{\prime\prime}$ pixels.  Contours show Stokes $I$ emission, marking 2\% and 20\% of the maximum Stokes $I$ value in the globule.}
    \label{fig:q}
\end{figure}

\begin{figure}
    \centering
    \includegraphics[width=0.47\textwidth]{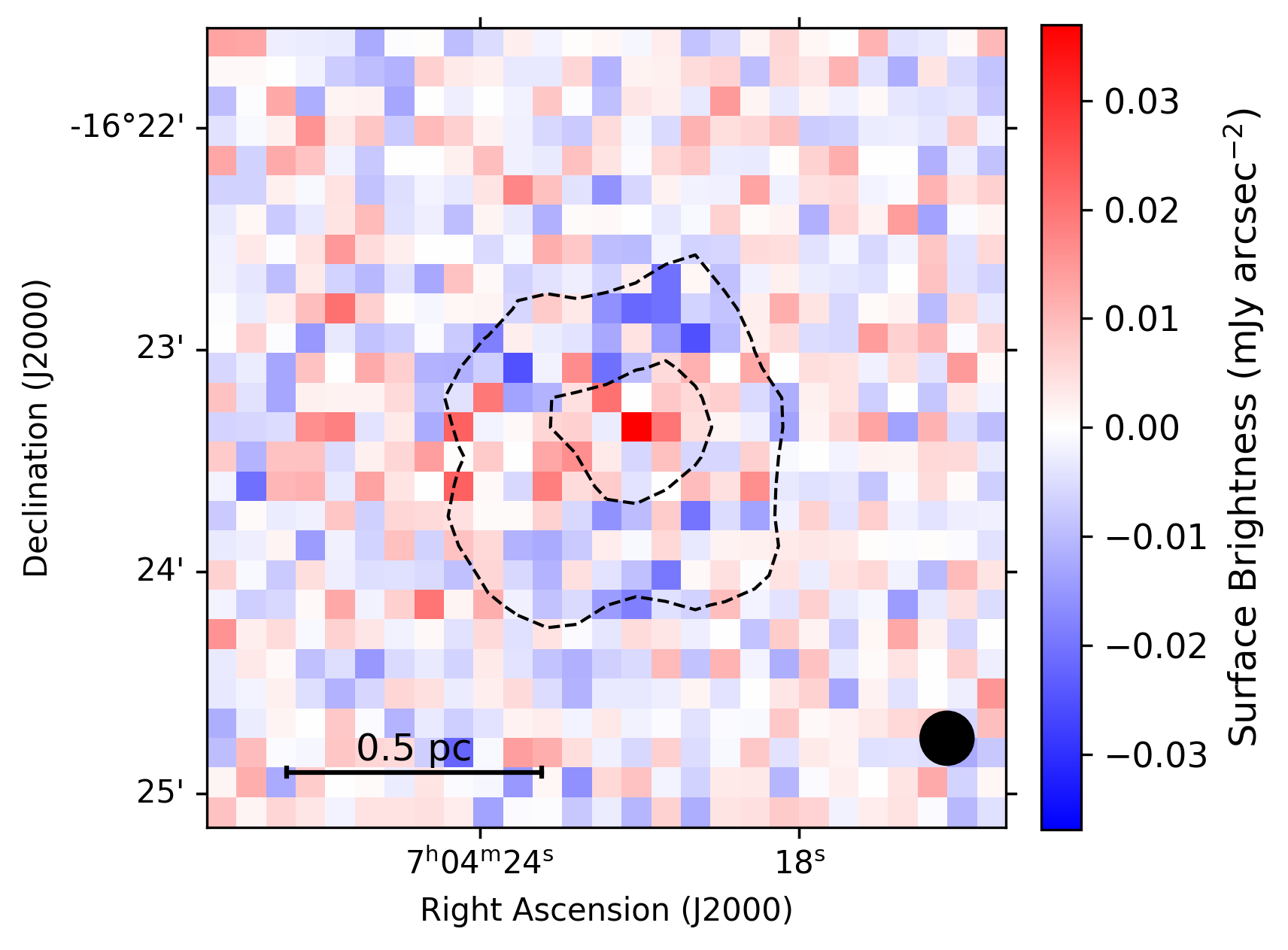}
    \caption{Stokes U emission in CB 54, shown on 8$^{\prime\prime}$ pixels.  Contours show Stokes $I$ emission, marking 2\% and 20\% of the maximum Stokes $I$ value in the globule.}
    \label{fig:u}
\end{figure}

\begin{figure}
    \centering
    \includegraphics[width=0.47\textwidth]{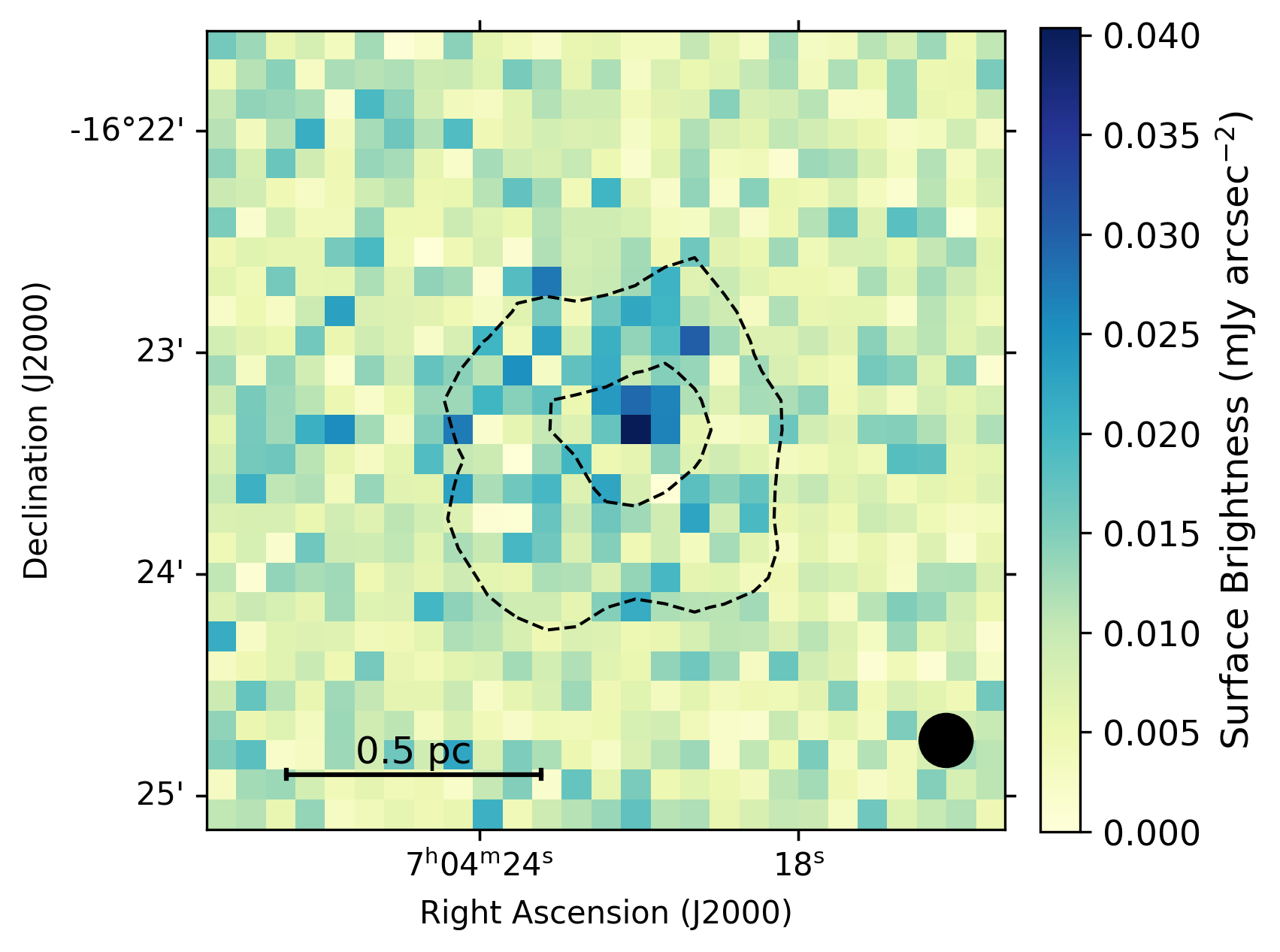}
    \caption{Non-debiased polarised intensity PI$^{\prime}$ emission in CB 54, shown on 8$^{\prime\prime}$ pixels.  Contours show Stokes $I$ emission, marking 2\% and 20\% of the maximum Stokes $I$ value in the globule.}
    \label{fig:pi}
\end{figure}

In this appendix we present maps of Stokes $Q$ (Figure~\ref{fig:q}) and $U$ (Figure~\ref{fig:u}) emission, and of non-debiased polarised intensity $PI^{\prime}$ (Figure~\ref{fig:pi}), on 8$^{\prime\prime}$ pixels.

%


\bsp	
\label{lastpage}
\end{document}